\documentclass[pre]{revtex4}
\usepackage[cp1251]{inputenc}
\usepackage[english,russian]{babel}

\usepackage{graphicx}
\usepackage{dcolumn}
\usepackage{eucal}
\usepackage[dvips]{epsfig}
\usepackage{amssymb}
\usepackage{amsmath}
\usepackage{color}

\begin{document}


\newcommand{\bs}{\boldsymbol}
\newcommand{\mbb}{\mathbb}
\newcommand{\mcal}{\mathcal}
\newcommand{\mfr}{\mathfrak}
\newcommand{\mrm}{\mathrm}

\newcommand{\ovl}{\overline}
\newcommand{\p}{\partial}

\renewcommand{\d}{\mrm{d}}
\newcommand{\lap}{\triangle}

\newcommand{\lan}{\bigl\langle}
\newcommand{\ran}{\bigl\rangle}

\newcommand{\bse}{\begin{subequations}}
\newcommand{\ese}{\end{subequations}}

\newcommand{\be}{\begin{eqnarray}}
\newcommand{\ee}{\end{eqnarray}}

\newcommand{\ga}{\alpha}
\newcommand{\gb}{\beta}
\newcommand{\gc}{\gamma}
\newcommand{\gd}{\delta}
\newcommand{\gr}{\rho}
\newcommand{\eps}{\epsilon}
\newcommand{\veps}{\varepsilon}
\newcommand{\gs}{\sigma}
\newcommand{\gf}{\varphi}
\newcommand{\go}{\omega}
\newcommand{\gl}{\lambda}

\renewcommand{\l}{\left}
\renewcommand{\r}{\right}

\title{\bf Theory of degenerate Bose gas without anomalous
averages}
\author{V.B. Bobrov, S.A. Trigger}
\address{Joint\, Institute\, for\, High\, Temperatures, Russian\, Academy\,
of\, Sciences, 13/19, Izhorskaia Str., Moscow\, 125412, Russia;\\
email:\,satron@mail.ru}

\begin{abstract}
Theory of a weakly non-ideal Bose gas in the canonical ensemble is
developed without assumption of the C-number representation of the
creation and annihilation operators with zero momentum. It is
shown that the pole of the "density-density"\, Green function
exactly coincides with the Bogolybov's phonon-roton spectrum of
excitations. At the same time in the one-particle excitation
spectrum a gap exists. The value of this gap is connected with the
density of particles in the "condensate".\\

PACS number(s): 05.30.Jp, 03.75.Kk, 03.75.Nt, 05.70.Fh\\

\end{abstract}

\maketitle

\section{Introduction}

1. Starting with the Bogolyubov's papers [1,2], the microscopic
theory of the degenerate Bose gas has been based on the special
assumption that the creation $a_0^{+}$ and annihilation $a_0$
operators of particles with zero momentum can be replaced by a
C-number
\begin{eqnarray}
a_0^{+}=a_0= (<N_0>)^{1/2}, \label{C1}
\end{eqnarray}
where $<N_0>$ is the average quantity of particles in the state
with ${\bf p}=0$ ("condensate").

This assumption leads to necessity of introducing in the theory
the \emph{anomalous averages} ("quasi-averages"), which is
unnatural for the homogeneous and isotropic system under
consideration. Though the Bogolyubov theory (see, e.g., [3,4])
gives rise to many important and widely recognized results, such
as the expression for the spectrum of excitations, explanation of
the experimental data and agreement with the Landau superfluidity
condition, there are serious doubts on the validity of the
relation as well as on the agreement of the Hamiltonian
corresponding to the assumption (\ref{C1}) with the original
Hamiltonian of the system under consideration [5,6].

In this work we suggest a selfconsistent\, noncontradictory
description of the degenerate Bose gas in which we do not use the
assumption (\ref{C1}). In this way we show, in particular, that
the gap in the spectrum of the one-particle excitations exists.
The statement that the spectrum of one-particle excitations has a
gap in parallel with the usual phonon-roton branch of the
collective excitations in degenerate Bose gas has been formulated
in [7].  Later the existence of a gap has been suggested in [8].
Spectra, thermodynamics and dynamical structure factor for weakly
non-ideal Bose gas below the condensation temperature $T<T_0$ have
been considered in [9-11] in terms of the dielectric formalism,
similar to one in plasma-like systems. The dielectric formalism on
the basis of the Bogolyubov assumption (\ref{C1}) for the
operators $a_0^{+}$ and $a_0$ was developed in [12]. Recently the
gap-existence problem for a weakly non-ideal Bose gas has been
considered by different methods in papers [13-16]. In fact, the
paper [14] confirmed (independently) and developed not only the
statement of Ref. [7] about existence of a one-particle gap in the
weakly non-ideal Bose gas below the condensation temperature, but
also the results of the papers [9-11] on a possibility of
considering the superfluid system without symmetry breaking. Of
course, the existence of a gap disagrees with a standard opinion
that the gap is missing [3]. But we do believe that the conclusion
about a missing gap is related to the Bogolubov simplifying
assumption (\ref{C1}), and is not obliged to persist in a more
rigorous theory.

It is necessary to stress that in the early papers both Landau and
Bogolyubov admitted that the one-particle excitation spectrum can
have a gap, but later they dropped this idea, because of
contradiction with observations of the phonon-roton branch of
spectrum at small $q$ in the neutron scattering experiments. The
approach which includes the both spectra, with and without a gap,
has not been found and coexistence of these two branches of
excitations was not suggested.

We suggest that in the limit of strong degeneration $T\rightarrow
0$, where $T$ is the temperature of the system, all the particles
tend to occupy the zero-momentum state, which means that
\begin{eqnarray}
\lim_{T\rightarrow 0}<N_0>=<N>, \label{C2}
\end{eqnarray}
where $<N>$ is the average number of particles in the system. This
statement is confirmed in the paper by the self-consistent
consideration. Below we use the canonical ensemble where $N$ is a
given C-number.

\section{Statistical sum and averages in canonical ensemble}

Let us consider the Hamiltonian of a non-ideal Bose gas of
particles with zero spin and a mass $m$ in a volume $V$
\begin{eqnarray}
H=\sum_p \varepsilon_{\bf p}\, a^+_{{\bf p}} a_{{\bf
p}}+\frac{1}{2V}\sum_{{\bf q, p_1,p_2}} u(q)a^+_{{\bf
p_1-q/2}}a^+_{{\bf p_2+q/2}}a_{{\bf p_2-q/2}}a_{{\bf p_1+q/2}},
\label{C3}
\end{eqnarray}
where $a^+_{p}$ and $a_{p}$ are the creation and annihilation
operators of particles with the momentum $\hbar{\bf p}$,
\begin{eqnarray}
[a_{{\bf p_2}}, a^+_{{\bf p_1}}]=a_{{\bf p_2}} a^+_{{\bf
p_1}}-a^+_{{\bf p_1}} a_{{\bf p_2}}=\delta_{{\bf p_1},{\bf p_2}},
\label{C4}
\end{eqnarray}
$\varepsilon_{\bf p}=\hbar^2{\bf p}^2/2m$ is the energetic
spectrum of a free particle and $u(q)$ is the Fourier-component of
the inter-particle interaction potential.

It is convenient to extract the particular term $U_0$ with ${\bf
q}=0$ from the sum over ${\bf q}$ in the Hamiltonian (\ref{C3}).
This term can be written as
\begin{eqnarray}
U_0=u(0)\frac{N(N-1)}{2V}, \,\; N=\sum_p a^+_{{\bf p}} a_{{\bf
p}}, \;\,\;u(0)=u(q=0)=u(q\rightarrow 0),\label{C5}
\end{eqnarray}
where $N$  is the operator of total number of particles, and $N$
is a C-number $N=<N>$. Here and below the brackets $<...>$ mean
the canonical-ensemble averaging. Since $U_0$ is also the C-number
one can write the statistical sum of the system under
consideration in the form
\begin{eqnarray}
Z=Sp\, \exp(-H/T)=\exp\left(-u(0)\frac{N(N-1)}{2VT}\right)Sp\,
\exp(-H_0/T),\label{C6}
\end{eqnarray}
where
\begin{eqnarray}
H_0=H-U_0=\sum_p \varepsilon_p\, a^+_{{\bf p}} a_{{\bf
p}}+\frac{1}{2V}\sum_{{\bf q\neq 0, p_1,p_2}} u(q)a^+_{{\bf
p_1-q/2}}a^+_{{\bf p_2+q/2}}a_{{\bf p_2-q/2}}a_{{\bf p_1+q/2}},
\label{C7}
\end{eqnarray}
Therefore, to provide convergence of the statistical sum
(\ref{C7}) in the thermodynamic limit $V\rightarrow \infty$ $N
\rightarrow \infty$, $n=N/V=const.$, the known condition $u(0)> 0$
has to be fulfilled. As $U_0$ is a C-number, it does not affect
any averaging at all, and for calculation of average values the
Hamiltonian $H$ is equivalent to $H_0$. For an arbitrary operator
$<A>$ we get
\begin{eqnarray}
<A>=Z^{-1} Sp\, \{\exp(-H/T)A \}=Z_0^{-1}Sp \,\{\exp(-H_0/T)A
\}\equiv <A>_0; \;\,\;  Z_0=Sp\, \exp(-H_0/T)\label{C8}
\end{eqnarray}

Let us now consider a more complicated situation, connected with
calculation of a time dependent correlation functions $f(t)$ of
the type
\begin{eqnarray}
f(t)=<[A(t), B(0)]>, \;\;\; A(t)=\exp(i H t/\hbar)A\exp(-i H
t/\hbar). \label{C9}
\end{eqnarray}
The time dependence of the operators $a^+_p(t)$ and $a_p(t)$ can
be represented in the form
\begin{eqnarray}
a^+_p(t)=\exp(iH_0 t/\hbar) a^+_p \exp(-iH_0t/\hbar)\exp(i N u (0)
t/V \hbar). \label{C10}
\end{eqnarray}
\begin{eqnarray}
a_p(t)=\exp(-i N u(0)t/V \hbar) \exp(iH_0 t/\hbar) a_p
\exp(-iH_0t/\hbar). \label{C11}
\end{eqnarray}
Therefore, if in each of the operators $A$ and $B$ the amounts of
creation and annihilation operators coincide (which is typical for
the operators of the physical variables) the time-dependent
correlation function can be written as
\begin{eqnarray}
f(t)=<[A(t), B(0)]>_0, \;\;\; A(t)=\exp(i H_0 t/\hbar)A\exp(-i H_0
t/\hbar). \label{C12}
\end{eqnarray}
On this basis below, in the framework of the canonical ensemble,
we consider the average values with the Hamiltonian $H_0$
(\ref{C7}). The free energy of the initial system with the
Hamiltonian $H$ (\ref{C3}), according to (\ref{C5})-(\ref{C7})
reads
\begin{eqnarray}
F=-T \ln Z=U_0+F_0, \;\;\; F_0=-T \ln Z_0. \label{C13}
\end{eqnarray}

\section{Equations for "density-density"\, Green function}

Experimentally the spectrum of collective excitations is found
usually from data on the well observable maxima in the dynamical
structure factor $S({\bf q},\omega)$ for ${\bf q} \neq 0$,
\begin{eqnarray}
S({\bf q},\omega)=\frac{1}{V}\int^\infty_{-\infty} \,exp(i\omega
t)<\rho_{{\bf q}} (t)\rho_{-{\bf q}} (0)>_0 dt, \label{C14}
\end{eqnarray}
\begin{eqnarray}
\rho_{{\bf q}} (t)= \sum_p \, a^{+}_{{\bf p-q}/2}(t)a_{{\bf
p+q}/2}(t), \label{C15}
\end{eqnarray}
where $\rho_{{\bf q}} (t)$ is the Fourier-component of the
operator of particle density in the Heisenberg representation. The
dynamical structure factor $S({{\bf q}},\omega)$ (\ref{C14}) is
directly connected [17] with the retarded density-density Green
function $\chi^R({{\bf q}},z)$ which is analytical in the upper
semi-plane of the complex variable $z$ ($Im z>0$),
\begin{eqnarray}
S({{\bf q}},\omega)=-\frac{2\hbar}{1- \exp(-\hbar \omega/T)}\,Im
\chi^R({{\bf q}},\omega+i0), \label{C16}
\end{eqnarray}
\begin{eqnarray}
\chi^R({{\bf q}},z)=-\frac{i}{\hbar V}\int^\infty_0 dt \,\exp(i z
t)<[ \rho_{{\bf q}} (t)\rho_{-{\bf q}}
(0)]>_0=\frac{1}{V}<<\rho_{{\bf q}} \mid \rho_{-{\bf q}}>>_{z},
\label{C17}
\end{eqnarray}
The definitions (\ref{C16}),(\ref{C17}) have to be taken in the
thermodynamic limit, where
\begin{eqnarray}
\lim_{T\rightarrow 0} <N_0>_0=<N>_0=N. \label{C18}
\end{eqnarray}

According to Eq.~(\ref{C14}) the function $\chi(q,z)$ can be
represented in the form
\begin{eqnarray}
\chi({{\bf q}},z) = \frac{1}{V} \,\sum_p \,F({\bf p,q},z),\,\;
F({\bf p,q},z)= \langle\langle \, a^{+}_{{\bf p-q}/2}a_{{\bf
p+q}/2} \mid \rho_{-q} \rangle\rangle_{z} \label{C19}
\end{eqnarray}
The equation of motion for the function $F({\bf p,q},z)$ with the
Hamiltonian $H_0$, determined by (Eq.~(\ref{C7})), can be written
in the form
\begin{eqnarray}
\left(\hbar z +\varepsilon_{\bf{p-q}/2}-\varepsilon_{\bf{p+q}/2}\right)\,F({\bf p,q},z)= f_{{\bf p-q}/2}-f_{{\bf p+q}/2}-\nonumber\\
\frac{1}{V}\sum_k u(k)\sum_{p_1}\langle\langle \, (a^{+}_{{\bf
p+k-q}/2}a^{+}_{{\bf p_1-k}/2}a_{{\bf p_1+k/2}}a_{{\bf p+q}/2}-
a^{+}_{{\bf p-q}/2}a^{+}_{{\bf p_1-k}/2}a_{{\bf p_1+k}/2}a_{{\bf
p-k+q}/2} )\mid \rho_{-q} \rangle\rangle_{z} \label{C20}
\end{eqnarray}
Here $f_{\bf p}$ is the one-particle distribution on the momenta
$\hbar {\bf p}$ function
\begin{eqnarray}
f_{\bf p}=<a^{+}_{{\bf p}}a_{{\bf p}}>_0 \label{C21}
\end{eqnarray}
For a temperature $T<T_0$, where $T_0$ is the temperature of
condensation the one-particle distribution function $f_{\bf p}$
can be represented as [18]
\begin{eqnarray}
f_{{\bf p}}=<N_0>\delta_{{\bf p},0}+f^T_{{\bf p}}(1-\delta_{{\bf
p},0}), \label{C22}
\end{eqnarray}
where \;$N_0=a_0^{+}a_0$ is the operator of the quantity of
particles with the momentum equal zero ("condensate")\;,
\;\;\;\;$f^T_{{\bf p}}=<a^{+}_{{\bf p}}a_{{\bf p}}>_0$ is the
one-particle distribution function with non-zero momenta (the
"overcondensate"\, states). Therefore,
\begin{eqnarray}
n=n_0 +\frac {1}{V} \sum_{p \neq 0} f^T_{{\bf p}}=n_0 + \int
\frac{d^3 p}{(2\pi)^3} f^T_{{\bf p}}, \label{C23}
\end{eqnarray}
where $n_0=<N_0>/V$ is the average density of particles in the
condensate. From (\ref{C20}) and (\ref{C22}) we find that the
function $F({\bf p,q},z)$ has singularities at ${\bf p}=\pm {\bf
q}/2$. Therefore, the density-density function \, $\chi(q,z)$
(\ref{C19}) can be represented in the form
\begin{eqnarray}
\chi(q,z) = \frac{1}{V}\,F({\bf q}/2,{\bf
q},z)\,+\frac{1}{V}\,F(-{\bf q}/2,{\bf q},z)\,+
\frac{1}{V}\sum_{p\neq\pm q/2}\,F^T({\bf p,q},z)\label{C24}
\end{eqnarray}
The index $T$ means, that the respective function describes the
"overcondensate"\, particles. The singularities (conditioned by
the condensate) in this function are absent. Then in the last term
in Eq.~(\ref{C24}) we can change summation by integration over
momenta. The functions $F(\pm {\bf q/2,q},z)$, extracted above,
satisfy, according to Eqs.~(\ref{C20}),(\ref{C22}), to the exact
equations of motion
\begin{eqnarray}
\left(\hbar z -\varepsilon_q\right)\,F({\bf q}/2,{\bf
q},z)=[\langle N_0 \rangle-f^T_{{\bf q}}]-\frac{1}{V}\sum_{k\neq
0} u(k)\sum_{p_1}\langle\langle \, (a^{+}_{\bf k} a^{+}_{{\bf p_1}-{\bf k}/2}a_{{\bf p_1}+{\bf k}/2}a_{\bf q} -\nonumber\\
a^{+}_0 a^{+}_{{\bf p_1}-{\bf k}/2}a_{{\bf p_1}+{\bf k}/2}a_{{\bf
q-k}})\mid \rho_{-q} \rangle\rangle_{z} \label{C25}
\end{eqnarray}
\begin{eqnarray}
\left(\hbar z+\varepsilon_q\right)\,F(-{\bf q}/2,{\bf
q},z)=-[\langle N_0 \rangle-f^T_{{\bf q}}]-\frac{1}{V}\sum_{k\neq
0}u(k)\sum_{p_1}\langle\langle \, (a^{+}_{{\bf k-q}} a^{+}_{{\bf p_1}-{\bf k}/2}a_{{\bf p_1}+{\bf k}/2}a_0 -\nonumber\\
a^{+}_{-{\bf q}} a^{+}_{{\bf p_1}-{\bf k}/2}a_{{\bf p_1}+{\bf
k}/2}a_{-{\bf k}})\mid \rho_{-q} \rangle\rangle_{z}. \label{C26}
\end{eqnarray}
Let us consider further the case of strongly degenerate gas, where
$T\rightarrow 0$. To find for this case the main terms, following
to the Bogolyubov's procedure, let us extract in the right sides
of Eqs.~(\ref{C25}),(\ref{C26}) the terms, which are determined by
the maximum quantity of the operators $a^{+}_0$ and $a_0$. In the
limit of strong degeneration, taking into account(\ref{C18}), we
can omit the other terms in the course of calculation of $F$.
Then, since $f^T_{{\bf q}}=f^T_{-{\bf q}}$ (${\bf q}\neq 0$)
Eqs.~(\ref{C25}),(\ref{C26}) take the form
\begin{eqnarray}
\left(\hbar z \mp \varepsilon_q\right)\,F^{(0)}(\pm {\bf q}/2,{\bf
q},z)=\pm [\langle N_0 \rangle-f^T_{{\bf q}}]\pm \frac{1}{V} u(q)
\langle\langle \, (a^{+}_0 a^{+}_0 a_{\bf q} a_0 + a^{+}_0
a^{+}_{-{\bf q}}a_0 a_0)\mid \rho_{-q} \rangle\rangle_{z}
\label{C27}
\end{eqnarray}
Eqs. (\ref{C27}) are exact at low temperature, since they
determine the functions $F^{(0)}(\pm {\bf q}/2,{\bf q},z)$. For
calculation the Green functions in the right side of Eqs.
(\ref{C27}) it is necessary to make some approximations.

\section{Determination of the collective excitations}

It should be emphasized that the Bogolyubov's approach, in which
the operators $a^{+}_0$ and $a_0$ are considered as C-numbers
leads to violation of the exact relations (\ref{C25}),
(\ref{C26}). In this approach (\ref{C1}), the main term $F^{(0)}$
of the function $F$ has the form [19]
\begin{eqnarray}
F^{(0)}(\pm {\bf q}/2,{\bf q},z)= \langle N_0 \rangle\,
\langle\langle \,a_{\pm\bf q}\mid a^{+}_{\pm \bf q}
\rangle\rangle_{z} \label{C28}
\end{eqnarray}
Respectively, instead the exact equations for two-particle Green
functions (\ref{C25}), (\ref{C26}) we obtain the equation of
motion for one-particle Green functions form (\ref{C28}). As is
shown below, the equations of motion for the two-particle Green
functions without the approximation on the C-number representation
of the operators $a^{+}_0$ и $a_0$ are essentially different from
the equations for the one-particle distribution functions.
Therefore, for calculation of the Green functions in the right
part of (\ref{C27}), the assumption (\ref{C1}) cannot be used. At
the same time the idea on C-number approximation for some
operators is, itself, very attractive. Below we use this idea in
the variant alternative to the Bogolyubov assumption. According to
(\ref{C18}) it is natural to accept that for calculation of the
Green functions in the limit of strong degeneration $T \rightarrow
0$ the C-number approximation has to be applied not to the
operators $a^{+}_0$ и $a_0$, but to the operator of the number of
particles in the "condensate"\, $N_0$

\begin{eqnarray}
N_0= \langle N_0 \rangle. \label{C29}
\end{eqnarray}
In this case from (\ref{C27}) directly follows
\begin{eqnarray}
\left(\hbar z - \varepsilon_q\right)\,F^{(0)}( {\bf q}/2,{\bf
q},z)=[\langle N_0 \rangle-f^T_{{\bf q}}]+\frac{<N_0>}{V}
u(q)\{F^{(0)}({\bf q}/2,{\bf q},z)+ F^{(0)}(-{\bf q}/2,{\bf
q},z)\} \label{C30}
\end{eqnarray}

\begin{eqnarray}
\left(\hbar z + \varepsilon_q\right)\,F^{(0)}(- {\bf q}/2,{\bf
q},z)=- [\langle N_0 \rangle-f^T_{{\bf q}}]-\frac{<N_0>}{V}
u(q)\{F^{(0)}({\bf q}/2,{\bf q},z)+F^{(0)}(-{\bf q}/2,{\bf q},z)\}
\label{C31}
\end{eqnarray}

From Eqs.~(\ref{C30}),(\ref{C31}) one can find the solutions for
the functions $F(\pm q/2,q,z)$
\begin{eqnarray} F({\bf q}/2,{\bf
q},\hbar z)=\frac{[<N_0>-f^T_{{\bf q}}](\hbar
z+\varepsilon_q)}{(\hbar z)^2-(\hbar\omega(q))^2};\;\; F(-{\bf
q}/2,{\bf q},z)=-\frac{[<N_0>-f^T_{{\bf q}}](\hbar
z-\varepsilon_q)}{(\hbar z)^2-(\hbar\omega(q))^2}\label{C32}
\end{eqnarray}
\begin{eqnarray}
\hbar\omega(q)\equiv \sqrt {\varepsilon_q^2+2 n_0 u(q)
\varepsilon_q}\label{C33}
\end{eqnarray}
The relation (\ref{C33}) for the spectrum $\hbar\omega(q)$
corresponds exactly to the known Bogolyubov expression [1,2]. By
substituting to (\ref{C19}), and taking into account that for the
case of strong degeneration the contribution of the functions
$F^T({\bf p,q},z)$ is negligible, we obtain the expression for the
main term $\chi^{(0)}(q,z)$ of the "density-density"\, Green
function $\chi(q,z)$
\begin{eqnarray}
\chi^{(0)}(q,z) = \frac{2 n_0 \varepsilon_q}{(\hbar
z)^2-(\hbar\omega(q))^2}\left\{1-\frac{f_q^T}{<N_0>}\right\}\label{C34}
\end{eqnarray}
As it is well known, the singularities of the function
$\chi^R(q,z)$ determine the spectrum of collective excitations in
the system. Therefore, under the assumption about C-number
behavior of the operator $N_0$ we obtain the Bogolyubov's result
for the spectrum of the collective excitations in the degenerate
and weakly interacting Bose gas. However, the question on the term
$f_{{\bf q}}^T/<N_0>$ in the figure brackets of (\ref{C33}) still
exists. The problem is in the behavior of the function $f_{{\bf
q}}^{id}$ for the ideal Bose gas [20]
\begin{eqnarray}
f^{id}_{{\bf q}}=\left\{\exp\left(\frac{\varepsilon(
q)}{T}\right)-1\right\}^{-1} \label{C35}
\end{eqnarray}
In the limit of small wave vectors $q$ the function $f^{id}_{{\bf
q}}$ converges at non-zero temperatures (further in the text
$1/q^2$-divergence). Moreover
\begin{eqnarray}
\lim_{T\rightarrow 0}\lim_{q \rightarrow 0}f^{id}_q\neq
\lim_{q\rightarrow 0}\lim_{T \rightarrow 0}f^{id}_q \label{C36}
\end{eqnarray}
The similar problem arises when (\ref{C22}) is used.

\section{One-particle excitations, gap and the self-consistent distribution function}

To calculate the distribution function $f_q^T$ for the
"overcondensate"\, particles in Bose gas let us consider the
one-particle Green function $g^R(q,z)$
\begin{eqnarray}
g^{R}(q,z) = \langle\langle \, a_{\bf q} \mid a^{+}_{\bf q}
\rangle\rangle_{z},\;\; {\bf q} \neq 0\label{C37}
\end{eqnarray}
This function is directly connected with the distribution function
$f_q^T$ by the relation [21]
\begin{eqnarray}
f_q^T=\int_{-\infty}^\infty
\frac{d\omega}{2\pi}g^{<}(q,\omega),\;\; g^{<}(q,\omega)= -2\hbar
\left\{\exp\left(\frac{\hbar\omega}{T}\right)-1\right\}^{-1} Im\,
g^{R}(q,\omega+i0)\label{C38}
\end{eqnarray}
The equation of motion for the Green function $g^{R}(q,z)$ for $q
\neq 0$ reads
\begin{eqnarray}
\left(\hbar z - \varepsilon_q\right)\,g^{R}(q,z)=1+
\frac{1}{V}\sum_{k\neq 0} u(k)\sum_{p}\langle\langle \,
a^{+}_{{\bf p+k}} a_{{\bf p}}a_{{\bf q+k}}\mid a^{+}_{\bf q}
\rangle\rangle_{z}. \label{C39}
\end{eqnarray}
As for the "density-density"\, Green function, we consider the
case of a strong degeneration and extract in the right-hand side
of Eq.~(\ref{C39}) the terms with maximum quantity of the
operators $a^{+}_0$ and $a_0$.  Then from Eq.~(\ref{C39}) we find
\begin{eqnarray}
\left(\hbar z - \varepsilon_q\right)\,g^{R}(q,z)=1+\frac{1}{V}
u(p)\langle\langle \, a^{+}_0 a_{\bf q} a_0 \mid a^{+}_{\bf q}
\rangle\rangle_{z}(1-\delta_{q,\,0}), \label{C40}
\end{eqnarray}
Further, by using again the assumption about the C-number behavior
of the operator $N_0$, from Eq.~(\ref{C40}) we obtain
\begin{eqnarray}
g^{R}(q,z)=\frac{1}{\hbar z -E_q}, \label{C41}
\end{eqnarray}
The expression for the spectrum of the one-particle excitations
$E_q$ is given by
\begin{eqnarray}
E_q=\varepsilon_q+n_0 u (q), \label{C42}
\end{eqnarray}
From Eqs.~(\ref{C37}),(\ref{C40}),(\ref{C41}) for $T \ll T_0$ we
obtain
\begin{eqnarray}
f_q^T=\frac{1}{\exp(E_q/T)-1}. \label{C43}.
\end{eqnarray}
Therefore, the function $f_q^T$ is finite for $q \rightarrow 0$.
Moreover, in the limit of strong degeneration $T \rightarrow 0$
\begin{eqnarray}
f_q^T \rightarrow 0 \label{C44}
\end{eqnarray}
for arbitrary values of $q$, in contrast with the case of of the
ideal Bose gas. Therefore, the representation (\ref{C22}) for the
one-particle distribution function $f_p$ is valid, the initial
suggestions (\ref{C2}),(\ref{C18}) are satisfied and the used
procedure of extraction of the main terms is self-consistent.

As it follows from (\ref{C42}) in the spectrum of one-particle
excitations the appeared gap equals to
\begin{eqnarray}
\Delta=E_{q\rightarrow 0}=n_0 u(0) \label{C45}.
\end{eqnarray}
A value of the gap is determined by the density of particles in
the "condensate". The existence of the gap permits to extend
essentially the applicability of the results obtained for
$T\rightarrow 0$.  It is obvious, that in many applications the
condition $T\rightarrow 0$ is equivalent to the condition $T\ll
\Delta$.

Taking into account (\ref{C44}) we can rewrite Eq.(\ref{C34}) for
the function $\chi^{(0)}(q,z)$ in the form
\begin{eqnarray}
\chi^{(0)}(q,z) = \frac{2 n_0 \varepsilon_q}{(\hbar
z)^2-(\hbar\omega(q))^2}\label{C46}
\end{eqnarray}

On the basis of (\ref{C46}) practically all known results for the
thermodynamical functions of the degenerate Bose gas can be
reproduced, as it was done in [9-11].

Therefore, in contrast with the approach, based on the C-number
approximation for the operators $a_0^{+}$ and $a_0$, by use the
C-number approximation for the operator $N_0$ we find the spectra
of the collective and one-particle excitations are different. Both
spectra, as it's easy to see, satisfy the Landau condition for
superfluidity. For the one-particle spectrum the Landau condition
is satisfied for the transitions between the "condensate"\ and the
"overcondensate"\ state. The Landau condition is, naturally,
violated for transitions between the "overcondensate"\ states.

\section{Conclusions}

Summarizing the performed consideration we can assert that by
calculation of the Green functions for the highly degenerate Bose
gas on the basis of the C-number approximation for the operator
$N_0$:

A) The problem of $1/q^2$ divergence, which arises for the ideal
Bose gas, can be solved;

B) The system has two different branches of excitations - the
one-particle and the collective ones, both satisfying the Landau
condition of superfluidity;

C) The one-particle spectrum of excitations contains the gap in
the region of small wave vectors, connected with the existence of
the "condensate"\;;

D) The spectrum of the collective excitations corresponds to the
"phonon-roton"\ excitations observed in the experiments on the
inelastic neutron scattering;

E) The necessity of anomalous averages (quasi-averages) for the
description of the degenerate Bose-gas is absent.

Therefore, the inter-particle interaction in Bose-systems leads
not only to the drastic difference (in comparison with the ideal
Bose gas) in the structure of the collective excitations, which
are described by "density-density"\ Green function, but also to
the crucial change in the distribution function of the
one-particle excitations and in the one-particle spectrum of
excitations for the "overcodensate"\ particles.

On the basis of the obtained results the special diagram technique
can be developed, similar to [19], but with the use of the
C-number approximation for the operator $N_0$.

The principle difference between the results of this work and the
results of the "traditional"\ C-number approximation for the
operators $a_0^{+}$ and $a_0$ is the existence of the gap in the
spectrum of the one-particle excitations. The above analysis shows
that this gap cannot manifest itself in the experiments on
nonelastic neutron scattering in superfluid Helium [22,23].
However, such possibility cannot be excluded [16] in the
experiments on the Raman light scattering. Moreover, in [24],
where such experiments are described, there is a direct indication
of existence of the gap.

\section*{Acknowledgment}
The authors thank M.V. Fedorov, Yu. A. Kuharenko and A.G.
Zagorodny for the useful discussions. The authors express
gratitude to the Netherlands Organization for Scientific Research
(NWO) for support of their investigations on the problems of
statistical physics.

\end{document}